\documentclass[aps,pra,twocolumn,showpacs,groupedaddress]{revtex4}

\usepackage{graphicx,bm}
\usepackage{amsmath,amssymb}
\usepackage{hyperref}
\usepackage{wasysym}

\usepackage{graphicx,amsfonts,amsbsy}

\newcommand{\uy}{\hat{\mathbf{y}}}
\newcommand{\uz}{\hat{\mathbf{z}}}

\newcommand{\di}{d}

\newcommand{\Kv}{\mathbf{K}}
\newcommand{\kv}{\mathbf{k}}
\newcommand{\rv}{\mathbf{r}}
\newcommand{\cv}{\mathbf{c}}
\newcommand{\bv}{\mathbf{b}}

\newcommand{\vv}{\mathbf{v}}

\begin{document}

\title{Vortex-lattice pinning in two-component Bose-Einstein
condensates}

\author{M. P. Mink}
\email{m.p.mink@uu.nl}

\affiliation{Institute for Theoretical Physics, Utrecht
University, Leuvenlaan 4, 3584 CE Utrecht, The Netherlands}

\author{C. Morais Smith}

\affiliation{Institute for Theoretical Physics, Utrecht
University, Leuvenlaan 4, 3584 CE Utrecht, The Netherlands}

\author{R. A. Duine}

\affiliation{Institute for Theoretical Physics, Utrecht
University, Leuvenlaan 4, 3584 CE Utrecht, The Netherlands}
\date{\today}


\begin{abstract}
We investigate the vortex-lattice structure for single- and
two-component Bose-Einstein condensates in the presence of an
optical lattice, which acts as a pinning potential for the
vortices. The problem is considered in the mean-field quantum-Hall
regime, which is reached when the rotation frequency $\Omega$ of
the condensate in a radially symmetric trap approaches the
(radial) trapping frequency $\omega$ and the interactions between
the atoms are weak. We determine the vortex-lattice phase diagram
as a function of optical-lattice strength and geometry. In the
limit of strong pinning the vortices are always pinned at the
maxima of the optical-lattice potential, similar to the
slow-rotation case. At intermediate pinning strength, however, due
to the competition between interactions and pinning energy, a
structure arises for the two-component case where the vortices are
pinned on lines of minimal potential.

\end{abstract}
\vskip2pc

\maketitle
\section{Introduction}
A characteristic property of a superfluid is that it supports
angular momentum through quantized vortices
\cite{onsager49,feynman55}. If many vortices are present in the
system, they arrange themselves in a hexagonal (Abrikosov) lattice
\cite{abrikosov57}. The latter was observed experimentally in
type-II superconductors \cite{tsakadze78,yarmchuk79} and more
recently in a single-component Bose-Einstein condensate (BEC)
\cite{madison00,aboshaeer01}. In the case of rotating
two-component condensates, the ground-state vortex lattice can
have a non-hexagonal structure, since the vortices in both
components can move with respect to each other. It was predicted
that, among other structures, the vortex lattice in a
two-component condensate can have an interlaced square structure
\cite{mueller02}. This structure was indeed observed
experimentally \cite{schweikhard04}.

The Abrikosov lattice structure arises in a single-component BEC
due to vortex-vortex interactions. One can also apply an
optical-lattice potential to the condensate consisting of a
regular pattern of potential minima and maxima. The prediction
\cite{reijnders04,reijnders05a,pu05} that vortices are pinned at
the maxima of an optical-lattice potential for sufficiently large
strength was indeed confirmed experimentally \cite{tung06}. Other
theoretical work on rotating Bose gases in optical lattices
focuses on the system near the superfluid-Mott-insulator
transition \cite{wu04a,goldbaum08,lundh08ar} and on the effect of
incommensurability between the vortex lattice and the optical
lattice \cite{goldbaum08ar}.

A remaining challenge is to determine the phase diagram of a
rotating two-component condensate in an optical-lattice potential.
Previous theoretical work focused on the slow-rotation limit and
neglected the effect of the non-zero total particle density inside
the vortex core \cite{reijnders05a}. In the interlaced
vortex-structure in a two-component BEC, the vortex lattices do
not lie on top of each other but are displaced by a fixed offset.
This feature shows that the inter-component interaction inside the
vortex cores is important. Therefore, the approach from
Ref.~\cite{reijnders05a} is not suitable to determine all the
vortex structures in rotating two-component BEC's.

Here, we consider instead the mean-field Quantum Hall regime,
where the angular momentum of the condensate is so high that the
wave function resides in the lowest Landau level (LLL), but
mean-field theory remains valid \cite{ho01,cooper08ar}. In this
regime the wave function is completely determined by the positions
of the vortices and no further approximations are needed. We
extend the method from Refs.~\cite{ho01,mueller02} to calculate
the optical-lattice energy for a wave function in the LLL. The
result is used to determine the phase diagram of a
single-component condensate in an optical lattice of arbitrary
geometry and a two-component condensate in a square optical
lattice. For single-component condensates, we find phases in which
the vortices are pinned on lines of maximal potential and phases
in which they are pinned at the pinning centers, which is
consistent with previous theoretical and experimental results
\cite{abrikosov57,reijnders04,reijnders05a,pu05,tung06,aboshaeer01,madison00}.
In the two-component case, we find the interlaced square lattice
in the absence of pinning, and also a new phase where the vortices
are pinned on lines of {\it minimal} potential.

The remainder of this paper is organized as follows. In
Sec.~\ref{sec:enfun} we evaluate the energy functional, including
the optical lattice, of the system in the LLL regime. The result
is used in Sec.~\ref{sec:pd} to determine the vortex phase
diagrams of single- and two-component BEC's in an optical lattice.
In Sec.~\ref{sec:conc} we present our conclusions.

\section{Vortex Pinning in the LLL: Theory}
\label{sec:enfun} We consider a rotating two-component BEC in an
optical-lattice potential. In this section, we discuss the
condensate wave function, the single particle energies, and
interaction energies in the LLL regime (see also
Refs.~\cite{ho01,mueller02}). Next, we evaluate the contribution
of the optical-lattice potential to the energy of the system.

\subsection{Energy functional}
We are mainly interested in the two-dimensional (2D) ordering of
the vortices. Therefore, we assume that the condensate has a small
effective size $d_z$ in the $z$-direction and consider a 2D wave
function. At the mean-field level, a two-component BEC is
described by two macroscopic condensate wave functions
$\psi_1(\rv)$ and $\psi_2(\rv)$, where $\rv = (x,y)$. The
condensate rotates with an angular velocity $\Omega$ around the
$z$-axis, thus it is convenient to transform to a frame that
co-rotates with the condensate. The wave functions $\psi_j$
describing the condensate are found by minimization of the energy
functional in the rotating frame
\begin{equation} \label{eq:enfun}
\mathcal{K} = \sum_{j=1,2}\left[ \int \di \rv \ \psi_j^*(\rv)(h_j
- \Omega L_z ) \psi_j(\rv)\right] +\mathcal{V}_I +
\mathcal{V}_{\text{OL}},
\end{equation}
where $\psi_j$ is normalized to the number of particles of species
$j$, $N_j$. The single-particle Hamiltonians $h_j$ are given by
$h_j = -(\hbar^2/2 M_j) \bm{\nabla}^2 + M_j \omega^2 r^2 /2$, with
$M_j$ the mass of a particle of species $j$, $\bm{\nabla} =
(\partial_x,\partial_y)$, $r = |\rv|$, and $\omega$ is the
frequency of the magnetic trapping potential in the radial
direction, which we assume to be the same for particles of both
species. The angular momentum operator is $L_z = -i \hbar \uz
\cdot (\rv \times \bm{\nabla})$. The interaction energy
$\mathcal{V}_I$ in Eq.~(\ref{eq:enfun}) reads
\begin{multline}
\mathcal{V}_I  =  \frac{1}{d_z} \int  \di \rv \  \Bigl(
\frac{g_1}{2}|\psi_1(\rv)|^4 + \frac{g_2}{2} |\psi_2(\rv)|^4  \\
   +  g_{12} |\psi_1(\rv)|^2 |\psi_2(\rv)|^2 \Bigl),
\end{multline}
where the {\it intra}- and  {\it inter}-component interaction
strengths are given by $g_j = 4 \pi \hbar^2 a_j / M_j$ and $g_{12}
= 2 \pi \hbar^2 a_{12} / M_{12}$, respectively. Here, $M_{12}$ is
the reduced mass, and $a_j$ and $a_{12}$ denote the intra- and
inter-component scattering lengths, respectively. In this work we
consider only repulsive interaction: $g_1,g_2,g_{12} > 0$.  The
optical-lattice potential couples to both species in the same way
and its energy $\mathcal{V}_{\text{OL}}$ in Eq.~(\ref{eq:enfun})
is given by
\begin{equation} \label{eq:enol}
\mathcal{V}_{\text{OL}} =  \int  \di \rv \  V_{\text{OL}}(\rv)
(|\psi_1(\rv)|^2 + |\psi_2(\rv)|^2),
\end{equation}
where the optical-lattice potential is given by
\begin{equation} \label{eq:oppor}
V_{\text{OL}}(\rv) = V_0 \left[\,\cos \left(\kv_1 \cdot \rv
\right) + \cos \left(\kv_2 \cdot \rv \right) \right],
\end{equation}
with the $\kv_i$'s denoting the reciprocal optical-lattice
vectors, and $V_0$ the optical-lattice strength. The relation
between the real-space optical-lattice vectors $\bv_i$ and the
$\kv_i$'s is $\kv_1 = (2 \pi/A_{\text{OL}}) \bv_2 \times \uz$ and
$\kv_2 = - (2 \pi/A_{\text{OL}}) \bv_1 \times \uz$, where
$A_{\text{OL}}= |\bv_1 \times \bv_2|$ is the area of the unit cell
of the optical lattice. We restrict our analysis to optical
lattices which have a rhombus-shaped unit cell. The angle between
the lattice vectors is denoted by $\phi$.

\subsection{Lowest Landau level}
It is shown in Ref.~\cite{ho01} that in the absence of an optical
lattice ($V_0 =0$) and in the presence of weak interaction
($\mathcal{V}_I$ small) the system enters the LLL regime when
$\Omega \uparrow \omega$. This result can be easily extended to
the case where $\mathcal{V}_{\text{OL}}$ is small. More
specifically, the criterium is that interactions and the optical
lattice do not cause transitions to higher Landau level states,
i.e., it must hold that $g_j n_j,g_{12} \sqrt{n_1 n_2} \ll \hbar
\omega$ and $V_0 \ll \hbar \omega$, where $\hbar \omega$ is the
Landau level gap and $n_j$ is the density of species $j$
particles. In this regime, the macroscopic wave function $\psi_j$
is completely determined by the position of the vortices and is
given by
\begin{equation} \label{eq:lllwf}
\psi_j(\rv) = \lambda_j \prod_{\alpha} (z - \xi^j_\alpha)
e^{-|z|^2/2\ell_j^2},
\end{equation}
where $\ell_j = \sqrt{\hbar/M_j \omega}$ is the magnetic length,
$z=x+iy$ is the complex position coordinate, $\lambda_j$ is a
normalization constant, and $\{ \xi^j_\alpha \}$ are the complex
positions of the vortices in the condensate of species $j$.

We assume that the vortex positions in Eq.~(\ref{eq:lllwf}) form
an infinite regular 2D lattice, and we project the functionals
$\mathcal{V}_I$ and $\mathcal{V}_{\text{OL}}$ onto the space of
wave functions with this property. The vortex lattice in
condensate 1 is spanned by lattice vectors $\cv_i$, which are
parameterized as
$$
\cv_1 = \sqrt{\frac{A_{\text{VL}}}{p \,\sin \theta}} (1,0) \quad
\text{and} \quad  \cv_2 = \sqrt{p A_{\text{VL}} \sin \theta} (\cot
\theta,1),
$$
where $p =|\cv_2|/|\cv_1|$ is the ratio between the lengths, the
angle between the $\cv_i$'s is $\theta$, and $\cv_1$ lies along
the $x$-axis. Since, for a given lattice, the lattice vectors can
always be chosen such that $p\geq 1$ and $\pi/3 \leq \theta \leq
\pi/2$, we restrict $p$ and $\theta$ to these ranges. The area of
the vortex-lattice unit cell is $A_{\text{VL}} = |\cv_1 \times
\cv_2|$. The reciprocal lattice vectors are $\Kv_1 = (2
\pi/A_{\text{VL}}) \cv_2 \times \uz$ and $\Kv_2 = - (2
\pi/A_{\text{VL}}) \cv_1 \times \uz$, where $\uz$ is the unit
vector in the $z$-direction. Note that within our approach it is
not necessary that $A_{\text{OL}}=A_{\text{VL}}$. The vortex
positions in the condensate 1 are given by $\Xi  = \{m_i \cv_i +
\vv_0 \}$, where the $m_i$ are integers, repeated indices $i \in
\{1,2\}$ are summed over, and $\vv_0$ is the offset of the vortex
lattice in condensate 1 from the origin. Then, it holds for the
particle density that
$$
|\psi_1(\rv)|^2 = f(\rv - \vv_0) e^{-r^2 / \sigma_1^2} \quad
\text{with} \quad \frac{1}{\sigma_1^2} = \frac{1}{\ell_1^2} -
\frac{\pi}{A_{\text{VL}}},
$$
where $\sigma_1$ is the effective condensate size in the radial
direction and $f$ is a structure function that is zero at the
positions $\{m_i \cv_i\}$ and has the lattice periodicity: $f(\rv
+ m_i \cv_i) = f(\rv)$ \cite{ho01,mueller02}.

In this paper, we want to study the situation in which the vortex
lattices in both components are commensurate, i.e., in which they
have the same geometry and unit cell area. It is therefore natural
to assume that atoms from species 1 and 2 are similar and to
restrict our analysis to the case where $N_1 \simeq N_2$, $M_1
\simeq M_2$, and $g_1 \simeq g_2$. In the remainder, we drop the
subscripts for these quantities. Thus, we assume that the vortex
lattices in both components are the same, up to a constant offset
$\rv_0$: If the set of vortex positions in component 1 is $\Xi$,
the set of vortex positions in component 2 is $\Xi + \rv_0$. Then,
both components are described by the same structure function $f$,
have equal effective radial size $\sigma \equiv \sigma_1$, and
equal vortex lattice unit cell area $A_{\text{VL}}$. Note that in
contrast to the non-rotating case, the system will not
phase-separate when $g_{12} > \sqrt{g_1 g_2}$. The reason for this
is that the system is already effectively phase-separated (albeit
incomplete), when it has an interlaced vortex-lattice structure
\cite{mueller02}. The particle density in component 2 is given by
$$|\psi_2(\rv)|^2 = f(\rv-\vv_0-\rv_0) e^{-r^2 / \sigma^2}.
$$
When $\Omega$ is close to $\omega$, the density spreads out in the
radial direction, so that $\sigma^2/A_{\text{VL}} \gg 1$, and it
can be shown \cite{mueller02} that the energy functional reads
\begin{equation} \label{eq:lllenfun1}
\mathcal{K}
 =  \frac{ 2 N \hbar(\omega - \Omega)\sigma^2}{\ell^2} + \frac{g
N^2}{2 \pi \sigma^2 d_z} \left(I  + \tilde{g}_{12} I_{12} \right)
+ \mathcal{V}_{\text{OL}},
\end{equation}
where $\tilde{g}_{12} = g_{12}/g$. The first term in
Eq.~(\ref{eq:lllenfun1}) is the contribution from the single
particle Hamiltonians $h_j$. The quantities $I$ and $I_{12}$
describe the intra- and inter-component interaction energy,
respectively, and are given by
\begin{equation} \label{eq:lllenfun2}
I = \sum_{\Kv} \left| f_{\Kv} \right|^2 \quad \text{and} \quad
I_{12} = \sum_{\Kv} \left| f_{\Kv} \right|^2 \cos (\Kv \cdot
\rv_0),
\end{equation}
where the $f_{\Kv}$ are the Fourier coefficients of the function
$f(\rv)$ and are given by
\begin{equation} \label{eq:lllenfun3}
f_\Kv = (-1)^{m_1+m_2+m_1 m_2} e^{-A_{\text{VL}} \Kv^2/8\pi}
\end{equation}
for a reciprocal lattice vector $\Kv = m_1 \Kv_1 + m_2 \Kv_2$,
with $m_1$ and $m_2$ integers. In the next section we calculate
$\mathcal{V}_{\text{OL}}$.

\subsection{Optical-lattice energy}

Using the Fourier expansion of the function $f(\rv)$, we normalize
the particle densities $|\psi_j(\rv)|^2$ to $N$:
$$
|\psi_1(\rv)|^2 = \frac{N}{\pi \sigma^2} \sum_{\Kv}
\tilde{f}_{\Kv} e^{i \Kv \cdot (\rv - \vv_0)} e^{-r^2/\sigma^2}
\text{ and}
$$
$$
|\psi_2(\rv)|^2 = \frac{N}{\pi \sigma^2} \sum_{\Kv} \hat{f}_{\Kv}
e^{i \Kv \cdot (\rv - \vv_0 - \rv_0)} e^{-r^2/\sigma^2},
$$
where we defined $ \tilde{f}_{\Kv} = f_{\Kv}/\sum_{\Kv'} f_{\Kv'}
e^{-i \Kv' \cdot \vv_0} e^{-\sigma^2 \Kv'^2/4}$ and $\hat{f}_{\Kv}
= f_{\Kv}/\sum_{\Kv'} f_{\Kv'} e^{-i \Kv' \cdot (\rv_0 + \vv_0)}
e^{-\sigma^2 \Kv'^2/4}$. We use that when $\Omega \to \omega$, it
holds that $\sigma^2 / A_{\text{VL}} \gg 1$ and since $\Kv^2 \sim
1/A_{\text{VL}}$ if $\Kv \neq 0$, only those terms in the
summations in the denominators of $\tilde{f}_{\Kv}$ and
$\hat{f}_{\Kv}$ for which $\Kv = 0$ survive. Hence,
$\tilde{f}_{\Kv} = \hat{f}_{\Kv} = f_\Kv$. By substituting the
expressions for $|\psi_1(\rv)|^2$ and $|\psi_2(\rv)|^2$ into
Eq.~(\ref{eq:enol}) we find
$$
\mathcal{V}_{\text{OL}} =  \frac{N V_0}{2} \sum_{\Kv,j} f_\Kv
e^{-i \Kv \cdot \vv_0} \left(1 + e^{-i \Kv \cdot \rv_0} \right)
G_{j,\Kv}~,
$$
where
\begin{align*}
G_{j,\Kv} & =  \frac{2}{\pi \sigma^2} \int \di \rv \  e^{i \Kv
\cdot \rv} e^{-r^2/\sigma^2} \cos \left(\kv_j \cdot \rv \right)\\
&= \left(e^{ -(\Kv + \kv_j)^2 \sigma^2 /4} + e^{ -(\Kv - \kv_j)^2
\sigma^2 /4} \right).
\end{align*}
We again use that in the fast-rotating limit $(\Kv + \kv_j)
\sigma^2 \gg 1$ or $(\Kv - \kv_j) \sigma^2 \gg 1$ unless
$\Kv=-\kv_j$ or $\Kv=\kv_j$, respectively. Thus, the Gaussian
terms of $G_{j,\Kv}$ are very small unless their argument is zero.
Therefore, it is reasonable to approximate $G_{j,\Kv}$ by the sum
of two Kronecker delta's: $G_{j,\Kv} = \delta_{\Kv, \kv_j} +
\delta_{- \Kv, \kv_j}$. Using that $G_{j,\Kv}$ and $f_\Kv$ are
even in $\Kv$, we find that
\begin{multline} \label{eq:lllol}
\mathcal{V}_{\text{OL}} = N V_0 \sum_{\Kv,j} f_{\kv_j}
\delta_{\Kv, \kv_j} \{ \cos \left(\kv_j \cdot \vv_0 \right) \\
   + \cos \left[\kv_j \cdot (\vv_0 + \rv_0) \right] \}.
\end{multline}

From Eq.~(\ref{eq:lllol}) it follows that the system can only gain
pinning energy if there are reciprocal vortex-lattice vectors
$\Kv$ equal to $\kv_1$ and/or $\kv_2$. In that case the vortices
are pinned on equally spaced lines, as we show below. Assume for
concreteness that $\kv_1 = \Kv = m_1 \Kv_1 + m_2 \Kv_2$, with
$m_1$ and $m_2$ integers. Using the definitions of the reciprocal
optical-lattice and vortex-lattice vectors given above, this is
equivalent to $m_1 \cv_1 - m_2 \cv_2 =
(A_{\text{VL}}/A_{\text{OL}}) \bv_2$. Since the $\cv_i$'s are
linearly independent vectors, there is a pair of {\it real}
numbers $(r_1,r_2)$ such that $r_1 \cv_1 + r_2 \cv_2 = \bv_1$. By
taking outer products between the left- and right-hand sides of
the last two equalities we obtain $|m_1 r_2 + m_2 r_1| =1$. By
inverting the expressions for the $\bv_i$'s, we reach the
conclusion that the coefficient of $\bv_1$ in the expansion of
both $\cv_i$'s is an integer. Thus, the vortices are pinned on the
collection of lines $\{n \bv_1 + r \bv_2 + \vv_0,n \in \mathbb{Z}
\text{ and } r \in \mathbb{R}\}$. Analogously, we can show that if
there is a reciprocal vortex lattice vector $\Kv$ equal to $\kv_2$
the vortices are pinned on the lines $\{r \bv_1 + n \bv_2 +
\vv_0,n \in \mathbb{Z} \text{ and } r \in \mathbb{R}\}$. Since
vortices are density minima, one intuitively expects that the
lines are always lines of maximal potential and that $\vv_0 = 0$.
Nonetheless, it turns out that in the two-component case there are
regions in the phase diagram that have non-zero $\vv_0$.

In the case that there are reciprocal vortex-lattice vectors
$\Kv_a$ and $\Kv_b$, equal to $\kv_1$ and $\kv_2$, respectively,
the vortex positions lie at the intersection of the two
collections of lines we mentioned above, i.e., on the positions
$\{m_1 \bv_1 + m_2 \bv_2 + \vv_0,m_i \in \mathbb{Z} \}$.

\section{\label{sec:pd}Vortex Pinning in the LLL: Phase Diagrams}
In this section we determine phase diagrams of single- and
two-component BEC's using the results for the energy functional in
Eqs.~(\ref{eq:lllenfun1}-\ref{eq:lllenfun3}) and the
optical-lattice energy Eq.~(\ref{eq:lllol}). We consider first the
case of a single-component condensate in an optical lattice with
arbitrary unit-cell angle and then the case of a two-component
condensate in a square optical lattice. For simplicity, we
restrict our analysis to vortex lattices with one vortex per
optical-lattice unit-cell, i.e., to the case $A_{\text{VL}}=
A_{\text{OL}} \equiv A$. Then, the first term in
Eq.~(\ref{eq:lllenfun1}) is a constant, which we drop.

\subsection{Single-component lattices} For the single-component
case, the energy $\mathcal{K}_s$ of a condensate of $N$ particles
in an optical lattice is found from Eq.~(\ref{eq:lllenfun1}) by
setting $\tilde{g}_{12}$ and $\rv_0$ to zero and dividing all
terms by a factor 2
$$
\mathcal{K}_s
 =  \frac{gN^2}{4 \pi \sigma^2 d_z} I
+ N V_0 \sum_{\Kv,j} f_{\kv_j} \delta_{\Kv, \kv_j} \cos\left(\kv_j
\cdot \vv_0 \right).
$$
\begin{center}
\begin{figure}
\includegraphics[width = 0.45 \textwidth]{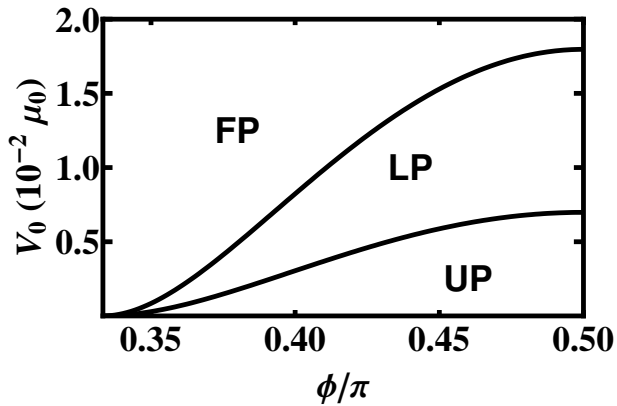}
\caption{\label{fig:1cpd}Phase diagram for a single component BEC
in an arbitrary optical-lattice potential. On the horizontal axis,
we show the angle $\phi$ of the optical lattice, from $\phi =
\pi/3$ (hexagonal lattice) to $\phi = \pi/2$ (square lattice). On
the vertical axis, we show the optical-lattice strength $V_0$ in
units of $\mu_0 \equiv gN/\pi\sigma^2 d_z$. The unpinned,
line-pinned, and fully-pinned phases are indicated by UP, LP, and
FP, respectively.}
\end{figure}
\end{center}
In Fig.~\ref{fig:1cpd} we display the phase diagram, which is
obtained by numerical minimization of $\mathcal{K}_s$ as a
function of the vortex lattice parameters for given optical
lattice angle $\phi$ and strength $V_0$. The latter is represented
in units of $\mu_0 \equiv gN/\pi \sigma^2 d_z$, which is equal to
the chemical potential $\mu$, up to a numerical factor of order 1.
The unpinned, line-pinned, and fully-pinned phases are indicated
by UP, LP, and FP, respectively. We describe these phases below.

In the {\it unpinned phase}, there is no $\Kv$ equal to $\kv_1$ or
$\kv_2$ and the last term of $\mathcal{K}_s$ is zero. The vortex
lattice ignores the optical lattice and the relative orientation
of the two lattices is not correlated. The incommensurability
between the vortex- and optical-lattice causes the optical-lattice
potential energy to average to zero.  The optimal unpinned vortex
lattice has $p=1$ and $\theta = \pi/3$. This corresponds to the
Abrikosov structure, and the result is consistent with
experimental and theoretical results obtained outside the LLL
regime \cite{madison00,aboshaeer01}.

\begin{center}
\begin{figure}
\includegraphics[width = 0.45 \textwidth]{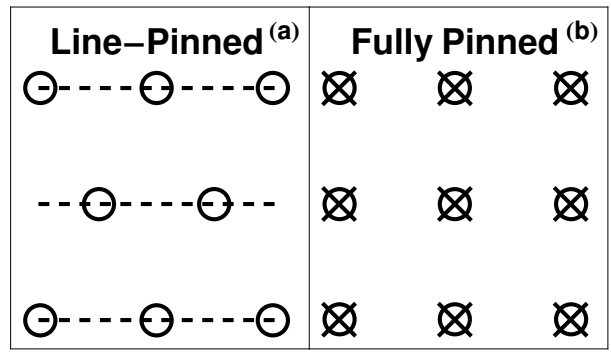}
\caption{\label{fig:1clat}The line-pinned and fully-pinned vortex
lattices for the case of an arbitrary optical-lattice angle $\phi$
and a square optical lattice, respectively. The dashed lines are
lines of maximal potential, the vortices are represented by
circles. In (a) the pinning centers (not shown) are located on the
dashed lines and in (b) they are represented by the crosses.}
\end{figure}
\end{center}

The {\it line-pinned phase} for an arbitrary optical-lattice angle
$\phi$ is shown in Fig.~\ref{fig:1clat}(a), where the $x$-axis is
in the direction of the dashed lines. In this phase the vortices
(circles) are pinned on the (dashed) lines of maximal potential,
specified by $\{r \bv_1 + n \bv_2 ,n \in \mathbb{Z} \text{ and } r
\in \mathbb{R}\}$. The pinning centers are not shown, but they are
located on the dashed lines. It holds that $\cv_1 = \bv_1$, $\vv_0
\cdot \uy = 0$, and the lattice vector $\cv_2$ is such that the
line-pinned vortex-lattice resembles the Abrikosov lattice
structure the closest: $\cv_2 = \bv_1/2 + (\bv_2 \cdot \uy) \uy$.
The energy does not depend on the $x$-coordinate of $\vv_0$: if we
shift the vortex lattice in the horizontal direction its energy
does not change. For the optical-lattice angle $\phi = \pi/3$, the
line-pinned vortex lattice has precisely the Abrikosov lattice
structure.

The {\it fully-pinned phase} is shown in Fig.~\ref{fig:1clat}(b)
for the square optical lattice. Here, all vortices are pinned on
the maxima of the optical lattice, which are represented by
crosses: $\cv_1 = \bv_1$, $\cv_2 = \bv_2$, and $\vv_0 = 0$. The
vortex lattice exactly matches the geometry of the optical
lattice.

If we start in Fig.~\ref{fig:1cpd} in the unpinned phase in the
case of a square optical lattice ($\phi = \pi/2$) and increase the
strength $V_0$, there is a transition to the line-pinned phase at
$V_0 = 0.007 \mu_0$. When we increase $V_0$ even further, we find
a second transition from the line-pinned to the fully-pinned phase
at $V_0 = 0.018 \mu_0$. The transitions are caused by the fact
that the system can gain pinning energy by transforming to a state
in which the vortices are pinned on the lines or points of maximal
potential. For optical-lattice angles $\phi < \pi/2$, one observes
that a smaller optical-lattice strength $V_0$ is sufficient to
trigger the transitions from the unpinned to the line-pinned and
from the line-pinned to the fully-pinned phase. At optical-lattice
angle $\phi = \pi/3$, it takes an infinitesimal $V_0$ to achieve
orientation locking. These observations can be explained by the
fact that for lower optical-lattice angles $\phi$ the fully-pinned
and line-pinned structures resemble more closely the hexagonal
Abrikosov lattice structure, which is the optimal structure in the
absence of pinning.

The orientation locking of the vortex lattice in an hexagonal
optical lattice $(\phi = \pi/3)$ was observed experimentally
\cite{tung06}. Contrary to our results, the authors find that a
non-zero minimum pinning strength is needed for orientation
locking, which they suggest is due to long equilibration times of
the system. Such non-equilibrium effects can not be investigated
with the equilibrium theory presented here. The fully-pinned phase
for a square optical lattice ($\phi= \pi/2$) was also found
earlier \cite{reijnders04,pu05,tung06}. The structural lattice
transitions in Fig.~\ref{fig:1cpd} typically occur for $V_0
\approx 0.01 \mu_0 \approx 0.01 \mu$, which is in rough agreement
with previous theoretical predictions \cite{reijnders04,pu05} and
experimental observations \cite{tung06}. Note that our line-pinned
phase is similar to the pinned phase found in
Ref.~\cite{reijnders04} for a {\it one-dimensional} optical
lattice, although the geometry of the optical lattice in our
system is {\it two-dimensional}. Furthermore, we note that the
half-pinned phase mentioned in
Refs~\cite{reijnders04,reijnders05a} is equivalent to the
line-pinned phase for $\phi = \pi/2$.

\subsection{Two-component lattices}

\begin{center}
\begin{figure}
\includegraphics[width = 0.47 \textwidth]{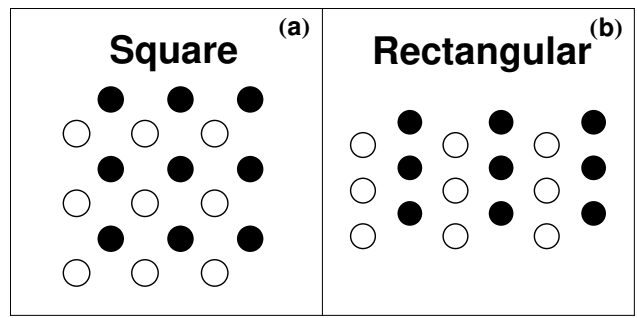}
\caption{\label{fig:2cnoOL} The interlaced square and rectangular
vortex lattices in a two-component condensate in the absence of
pinning. The empty (filled) circles correspond to vortices in
component 1 (2). For further details we refer the reader to
Ref.~\cite{mueller02}.}
\end{figure}
\end{center}

\begin{center}
\begin{figure}
\includegraphics[width = 0.45 \textwidth]{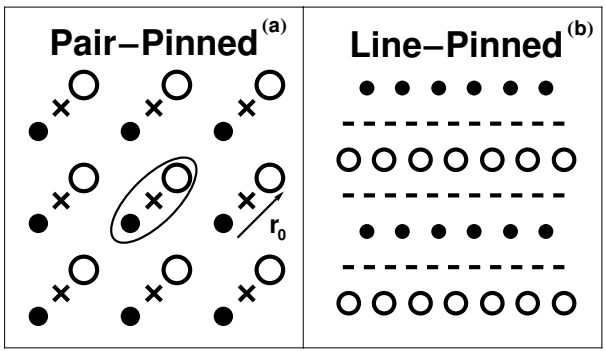}
\caption{\label{fig:2cpinlat}The pair-pinned and line-pinned
two-component vortex lattices. The crosses indicate the location
of the pinning centers, the empty circles the vortices of in
component 1, and the filled circles those in component 2. The
horizontal dashed lines are lines of maximal potential. In (b) the
pinning centers (not shown) are located on the dashed lines. }
\end{figure}
\end{center}

The energy functional $\mathcal{K}$ of a rotating two-component
condensate in an optical lattice is given by
Eqs.~(\ref{eq:lllenfun1}-\ref{eq:lllol}). 
In Ref.~\cite{mueller02}, the system was analyzed in the absence
of pinning, where $V_0=0$. The authors find that the structure of
the ground-state vortex-lattice depends on the value of
$\tilde{g}_{12}$. We show two possible configurations for $V_0 =
0$ in Fig.~\ref{fig:2cnoOL}, where the empty (filled) circles
correspond to vortices in component 1 (2). The interlaced square
structure, mentioned in the introduction, is the ground state when
$0.37 < \tilde{g}_{12} < 0.93$ (see Fig.~\ref{fig:2cnoOL}(a)). For
$\tilde{g}_{12} > 0.93$, the vortex lattice deforms into a
rectangular structure, as shown in Fig.~\ref{fig:2cnoOL}(b). The
ratio between the lengths of the two vortex lattice vectors $p$
continuously increases with increasing $\tilde{g}_{12}$. For
further details we refer the reader to Ref.~\cite{mueller02}.
\begin{center}
\begin{figure}
\includegraphics[width = 0.47 \textwidth]{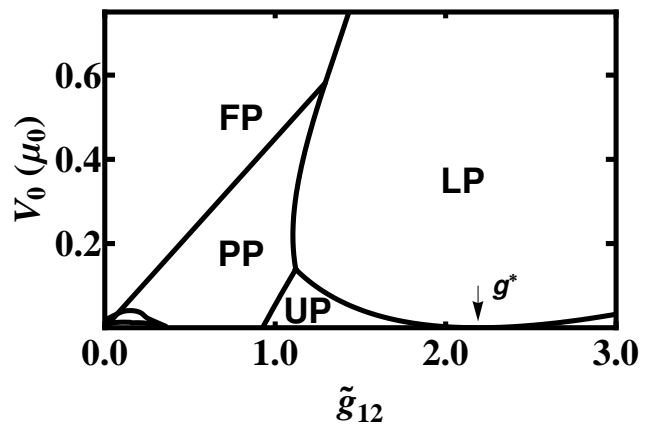}
\caption{\label{fig:2cpd}Vortex phase diagram of a two-component
condensate in the presence of a square optical lattice. On the
horizontal axis, we show the scaled inter-component interaction
strength $\tilde{g}_{12}$, and on the vertical axis, the
optical-lattice strength $V_0$ in units of $\mu_0$.}
\end{figure}
\end{center}
The vortex phase diagram of a two-component condensate in the
presence of a {\it square} optical lattice is shown in
Fig.~\ref{fig:2cpd}. On the horizontal axis we show the scaled
inter-component interaction strength $\tilde{g}_{12}$, and on the
vertical axis the optical-lattice strength $V_0$ in units of
$\mu_0$. First, we notice that the left below corner of the phase
diagram contains unpinned phases which are described in
Ref.~\cite{mueller02} and a line-pinned phase which is a trivial
extension to the two-component case of the single-component
line-pinned phase shown in Fig.~\ref{fig:1clat}(a). We will not
discuss these phases further. At larger values of $\tilde{g}_{12}$
and/or $V_0$ novel phases arise. In the pair-pinned (PP) structure
(see Fig.~\ref{fig:2cpinlat}(a)), the vortex lattice consists of
pairs of a component 1 and a component 2 vortex, which form a
square lattice. The vortices in a pair are displaced by $\rv_0 = r
(\cv_1 +  \cv_2)$ with respect to each other and the pairs are
pinned at the optical-lattice maxima in such a way that the
vortices in a pair lie at equidistant positions from their pinning
center. In Fig.~\ref{fig:2cpinlat}(a), we encircled one of these
vortex pairs for clarity. The value of $r$ at $V_0=0$ is 0.5, so
that the vortex lattice has the interlaced square structure shown
in Fig.~\ref{fig:2cnoOL}(a). When we increase $V_0$, the value of
$r$ decreases continuously from 0.5 to 0: The system gains pinning
energy if the vortices in a pair move closer together towards the
pinning center. When $r=0$ the system is in the fully-pinned phase
(FP), where the vortices in a pair are pinned on top of each other
at the potential maxima, analogous to the fully-pinned
single-component vortex lattice in Fig.~\ref{fig:1clat}(b).

At low values of $V_0$ and for $\tilde{g}_{12} \approx 1.2$ and
$\tilde{g}_{12} \approx 3$, the system is in the unpinned
interlaced rectangular structure (see Fig.~\ref{fig:2cnoOL}(b)),
which we denote by UP. At the value $\tilde{g}_{12} = 2.19 \equiv
g^*$, the ratio $p$ between lengths of the lattice vectors of the
unpinned rectangular lattice is equal to 4. Then, the rectangular
vortex lattice is commensurate with the square optical lattice and
can be line-pinned by an arbitrarily weak optical lattice. For
values of $\tilde{g}_{12}$ close to $g^*$, the transition to the
line-pinned phase occurs at small, but finite values of $V_0$. In
terms of lattice vectors, this phase is specified by $\cv_1 =
\bv_1 /2$, $\cv_2 = 2 \bv_2$, and $\rv_0 = (\cv_1 + \cv_2)/2$.
Interestingly, the lines on which the vortices in this phase are
pinned are not the lines of maximal potential, as one would expect
intuitively, but the lines of minimal potential, which is opposite
to the single-component case in Fig.~\ref{fig:1clat}(a). The
reason is that the {\it total} density is not minimal on the
(horizontal) lines on which the vortices lie, but halfway between
those lines. It follows from Eq.~(\ref{eq:lllwf}) that the healing
length of vortices in the LLL regime is equal to the distance
between the vortices and this causes the above-mentioned location
of the minima in the total density. The minima in the {\it total}
density {\it are} pinned on lines of maximal potential, as
expected.

\section{\label{sec:conc}Discussion and Conclusions}
In summary, we have considered vortex lattices in single- and
two-component BEC's in a 2D optical-lattice potential in the LLL
regime. We have incorporated the effects due to the optical
lattice and determined the phase diagram of a single-component
condensate in an optical-lattice potential with arbitrary
unit-cell angle and of a two-component condensate in a square
optical-lattice potential. For the single-component case, we find,
among others, phases that are pinned on lines of maximal potential
or at the potential maxima. In the two-component case, we find a
phase in which pairs of vortices are pinned at the potential
maxima, and a phase where vortices are pinned on the lines of {\it
minimal} potential. Note that Ref.~\cite{goldbaum08} also points
out the possibility of pinning at the minima of the
optical-lattice potential, albeit in a different regime than we
consider here.

As mentioned before, the criteria for the validity of the LLL
assumption are that the interaction energy per particle $gn$ and
the optical-lattice strength $V_0$ are much smaller than the LLL
gap $\hbar \omega$. Full pinning typically occurs at $V_0 = 0.01
\mu_0$ ($V_0 = 0.1 \mu_0$) for single-component (two-component)
condensates, so that at these typical values ($V_0 \ll \mu_0
\approx gn$) the LLL condition remains valid. This difference in
optical-lattice strength comes about because of the difference in
vortex filling factors considered. We don't expect fundamental
obstacles for observing vortex pinning in two-component
condensate, since the rotation velocities for which vortex pinning
in a single-component BEC and the interlaced square lattice
structure in a two-component BEC without pinning were observed are
similar \cite{schweikhard04,tung06}. Our results are consistent
with theoretical \cite{reijnders04,reijnders05a} and experimental
\cite{tung06} work for single-component condensates in an optical
lattice \cite{reijnders04,reijnders05a,tung06} and two-component
condensates without pinning \cite{schweikhard04}, which were
performed {\it outside} the LLL regime. Hence, we expect that our
results are, at least qualitatively, valid also outside the LLL
regime.

A possibility for further research is to relax the restriction
that there are one or two vortices per optical-lattice unit-cell
and study the effect of incommensurability between the vortex
lattice and the optical lattice. In particular, it is interesting
to investigate the situation in which the unit cell of the optical
lattice is smaller than the critical vortex-lattice unit-cell size
$\pi \ell^2$, which would require an extension of the formalism
presented here.

This work was supported by the Netherlands Organization for
Scientific Research (NWO) and by the European Research Council
under the Seventh Framework Program (FP7).

\end{document}